\documentclass[aps,pre,onecolumn,showpacs,showkeys,amsmath,amssymb,
groupedaddress]{revtex4}
\usepackage{graphicx}

\newcommand{\prt}{\partial}

\begin{document}

%\preprint{}

\title{Static and dynamic aspects of transonicity in Bondi accretion}

\author{Arnab K. Ray}
\email{akr@iucaa.ernet.in}
\affiliation{Inter--University Centre for Astronomy and Astrophysics \\
Post Bag 4, Ganeshkhind, Pune 411007, India}
\altaffiliation[Also at: ]{Harish--Chandra Research Institute, 
Chhatnag Road, Jhunsi, Allahabad 211019, India} 

\author{Jayanta K. Bhattacharjee}
\email{tpjkb@mahendra.iacs.res.in}
\affiliation{Department of Theoretical Physics \\
Indian Association for the Cultivation of Science \\
Jadavpur, Kolkata 700032, India}

\date{\today}

\begin{abstract}
Transonicity in a spherically symmetric accreting system has 
been considered in both the stationary and the dynamic regimes. 
The stationary flow, set up as a dynamical system, has been 
shown to be greatly unstable to even the minutest possible deviation 
in the boundary condition for transonicity. With the help of a
simple analytical model, and some numerical modelling, it has 
then been argued that the flow indeed becomes transonic and stable, 
when the evolution of the flow is followed through time. The 
time-dependent approach also shows that there is a remarkable 
closeness between an equation of motion for a perturbation in 
the flow, and the metric of an analog acoustic black hole. 
\end{abstract}



\maketitle

%\hrule
%\vspace{0.2cm}
%{\noindent \bf \small This article is an invited contribution 
%to a special issue of the Indian Journal of Physics, dedicated to 
%the revered memory of Prof. Amal Kumar Raychaudhuri.}
%\vspace{0.2cm}
%\hrule

\section{Introduction}
\label{sec1}

Accretion processes involve, in very simple terms, the flow dynamics
of astrophysical matter under the external gravitational influence
of a massive astrophysical object, like an ordinary star or a white 
dwarf or a neutron star~\cite{fkr92}.
The qualifier ``external" is to be stressed upon here, to
distinguish accretion processes from the self-gravity driven collapse of
a fluid system, as in the case of a star itself. The accreting astrophysical
matter whose fluid properties we are interested in, could be the 
interstellar medium --- as modelled by its spherically symmetric infall
on to an isolated accretor --- or stellar matter, as seen in a binary
system, where tidal deformation of a star, leads to matter flowing out
from it into the potential well of a compact companion~\cite{fkr92}.
In all of these cases, the fluid system is satisfactorily
described by a momentum balance equation (with gravity as an
external force), the continuity equation and a polytropic equation of
state.

In astrophysics, studies in accretion, in a formal sense, have been
carried out for more than half a century now. Initially, astrophysical
problems in the nature of what we understand to be accretion processes
at present, were studied by Hoyle and Lyttleton in the context of the
infall of matter on to a star moving through the interstellar
medium~\cite{bon52}.
In their methods, however, Hoyle and Lyttleton neglected the pressure
effects, with the argument that any heat generated would be radiated
away rapidly, so that the temperature of the infalling gas (and related
to that, 
the effects of pressure as well) would remain negligibly low~\cite{bon52}.
This was
found to be a satisfactory prescription for most cases of astrophysical
interest, which were then being studied. In 1952, however, in a very
important paper~\cite{bon52},
Bondi attacked this problem somewhat differently, by
taking into account the pressure effects. This work on spherically
symmetric infall of matter, on to a massive and attracting centre,
using formal
fluid dynamical equations, has assumed a paradigmatic status in accretion
studies~\cite{fkr92,skc90}.

Bondi studied the problem of spherical accretion in its stationary
limit, i.e. by only considering the extreme case of negligible dynamical
effects. In his own words the mathematical difficulties associated with
the problem of studying both the pressure effects (ignored by Hoyle and
Lyttleton) and the dynamical effects were ``insuperable"~\cite{bon52},
given the computational facilities available to him half a century ago.

The stationary equations would lead to various classes of solutions. Of
these the interesting ones would be those which obey the outer boundary
condition that at large radii the flow velocity would be highly subsonic,
i.e. small compared with the speed of sound, with the speed of sound itself
approaching a constant ``ambient" value at large radial distances. With
this outer boundary condition being satisfied, it would be physically
meaningful to consider only the entire class of solutions, which remain
subsonic everywhere, and the exceptional case of the lone transonic
solution. This exceptional solution, determined uniquely by the value of
the density at infinity, crosses the sonic point of the flow --- a point
which is given by the radius where the flow velocity smoothly matches
the speed of sound --- and acquires supersonic values at lesser radial
distances. Further, in one sense this transonic solution represents a
limiting value for the mass infall rate, for a given value of the density
at infinity, because any inflow rate higher than that given by the
transonic solution, will make the flow ``bounce" outwards~\cite{pso80}.

In this situation it became worthwhile to inquire into the natural
selection of a solution by a spherically symmetric accreting system.
Regarding this question Bondi himself offered some insights when he
observed in his paper that since it is with the transonic solution
that the lowest energy is associated, we may find a natural system in
this state~\cite{bon52}.
Furthermore, Bondi suggested that this result would also
be ``in agreement with the intuitive idea that, since there is nothing
to stop the process of accretion, it takes place at the greatest possible
rate"~\cite{bon52}, i.e. the rate given by the transonic solution.

These arguments by Bondi were essentially based on physical considerations.
His suggestion that a linear stability analysis might offer a clue was
also put to the test~\cite{pso80,gar79,td92}.
Subjecting the physically relevant stationary
inflow solutions to a linearised time-dependent perturbation led to the
result that all solutions displayed stable behaviour, and through this
method, no insight may be had as to the natural inclination of the
system for any particular solution. Garlick settled this point quite
categorically by stating that ``the plausible assumption that subsonic
flows are not realised in nature, and that critical flows always develop,
cannot be justified by a linear stability analysis, but rather by the
more fundamental arguments given by Bondi"~\cite{gar79}.

This then was the understanding that became firmly established --- that
the transonic solution enjoyed primacy over the subsonic ones. It was
easy to recognise that for black holes, this would be certainly true.
For one thing, instead of a physical surface, black holes have what is
known as an event horizon, which precludes all possibility of a pressure
build-up at smaller radii; for another, all matter reaching the event
horizon must do so supersonically, implying that it must display
transonic behaviour~\cite{nt73}.

The situation, however, was not so clear-cut if the accretor had a
hard surface like a neutron star or a white dwarf. For such an accretor,
it may be supposed that the accumulated matter would build up pressure
near the surface and cause the supersonic flow to be shocked down to
subsonic levels, although for a neutron star in particular, all
accreted matter is expected to be efficiently ``vacuum cleaned" away,
making it easier for the flow to remain supersonic~\cite{pso80}.

Then again there was the conceptual difficulty in understanding the
realisability of the transonic solution, by only an infinitely precise
determination of the outer boundary condition for the stationary
equations of the flow. An infinitesimal error in the determination of
the boundary condition would generate a solution far away from
transonicity. Within the framework of the stationary picture, it is
difficult to imagine that a natural physical system would be so precisely 
tuned.
On the other hand, quite intriguingly, this difficulty is resolved
when, instead of merely the stationary equations, the transonic
solution is tried to be numerically generated through a temporal
evolution (accounting for explicit time-dependence) of the spherically
symmetric accreting system~\cite{cos78,vit84,zmt96}.

In this article we make some attempt in addressing these issues. We
take up the stationary flow equations first and model them along the
lines of the equations governing a dynamical system. This establishes
the nature of the sonic point of the flow, as that of a saddle point.
Quite apart from the fact that a saddle point is inherently unstable,
among other adverse implications, this analysis illustrates why in the
stationary picture, after having started with an outer boundary condition,
so much difficulty is encountered in generating a solution that passes
through the sonic point~\cite{rb02}. The stationary transonic solution is
notoriously unstable under even the smallest of deviations from the
infinitely precise boundary condition that would be needed to generate
the solution.

Since in a realistic situation, an accreting system would have evolved
through only a finite span of time, we then consider the case of
explicit time-evolution of the flow solution(s), starting from a
reasonably physical initial condition (given at $t=0$). This gives us
to understand quite convincingly that not only would the transonic
solution be selected, but that the physical mechanism
that is very likely effective in making the selection is the one about
which Bondi has conjectured (entirely within the confines of the
stationary framework) --- that the flow solution with the lowest
specific energy associated with it, would be preferred to all the
others. It is important to note that this selection mechanism is
actually at work through the temporal evolution of the flow, and that
it is entirely non-perturbative in character~\cite{rb02}.

In keeping with this time-dependent approach to uphold the exclusive
selection of the transonic solution from a host of various  
solutions, we have finally carried out an 
interesting exercise, that is actually perturbative in nature. On 
imposing a linearised time-dependent perturbation on the constant 
matter flow rate, we 
have been able to identify a remarkable closeness between an equation
of motion for the perturbation, and the metric of an 
acoustic black hole. With the aid of this analogy we have argued that
contrary to common belief, even a perturbative treatment might be 
conveying a subtle hint in favour of transonicity.

\section{The equations of the spherically symmetric flow}
\label{sec2}

The variables that we need to consider are the velocity (radial
velocity only for the spherically symmetric flow), $v$, and the 
density, $\rho$. We ignore viscosity and write down the inviscid Euler  
equation for $v$ as
\begin{equation} 
\label{euler}
\frac{\partial v}{\partial t} + v \frac{\partial v}{\partial r} =
- \frac{1}{\rho} \frac{\partial P}{\partial r}
- \frac{\partial V(r)}{\partial r} , 
\end{equation}
where $P$ is the local pressure and $V$ is the potential due to
the gravity of the central accretor of mass $M$, given by 
$V(r) = -GM/r $. The pressure is related to the local density 
through a polytropic equation of state $P=K\rho^\gamma$,  
in which $K$ is a constant, and $\gamma$ is the polytropic 
exponent~\cite{sc39}, whose admissible range is given 
by $1<  \gamma < 5/3$, with this range having been restricted by the 
isothermal and the adiabatic limits, respectively. To know how 
$\rho$ evolves, we need the equation of continuity,
\begin{equation}
\label{con}
\frac{\partial \rho}{\partial t} + \frac{1}{r^2} \frac{\partial}
{\partial r} (\rho v r^2 ) = 0 . 
\end{equation}

Our system is specified by Eqs.(\ref{euler}) and (\ref{con}). We are 
interested in static solutions, the problem of which has been defined 
in Bondi's own words as follows --- ``A star of mass $M$ is at rest 
in an infinite cloud of gas, which at infinity is also at rest...The 
motion of the gas is spherically symmetrical and steady, the increase 
in the mass of the star being ignored so that the field of force is 
unchanging"~\cite{bon52}. Since transonic flows are our concern, we 
require that the static flow evolves from $v \longrightarrow 0$ as 
$r \longrightarrow {\infty}$ (the outer boundary condition) to 
$v> c_s (r)$ for small $r$, where $c_s(r)$ is the speed of sound 
given by 
$c_s^2 = \partial P/\partial \rho = \gamma K \rho ^{\gamma - 1}$. 
 
The stationary solution implies $\partial v/\partial t =
\partial \rho /\partial t = 0$, and hence we have $\rho \equiv
\rho (r)$ and $v \equiv v(r)$. This requirement renders 
Eqs.(\ref{euler}) and (\ref{con}), as
\begin{equation}
\label{stateuler}
v \frac{dv}{dr} + \frac{1}{\rho} \frac{dP}{dr} + \frac{GM}{r^2}=0
\end{equation}
and
\begin{equation}
\label{statcon}
\frac{1}{v} \frac{dv}{dr} + \frac{1}{\rho} \frac{d \rho}{dr}
+ \frac{2}{r} =0 , 
\end{equation}
respectively. It is to be noted here that the Eqs.(\ref{stateuler}) 
and (\ref{statcon}) remain invariant under the transformation 
$v \longrightarrow -v$, i.e. the 
mathematical problem for inflows ($v<0$) and outflows ($v>0$) is the 
same~\cite{arc99} in the stationary state. 

\begin{figure}[b]
\begin{center}
\includegraphics[scale=0.4]{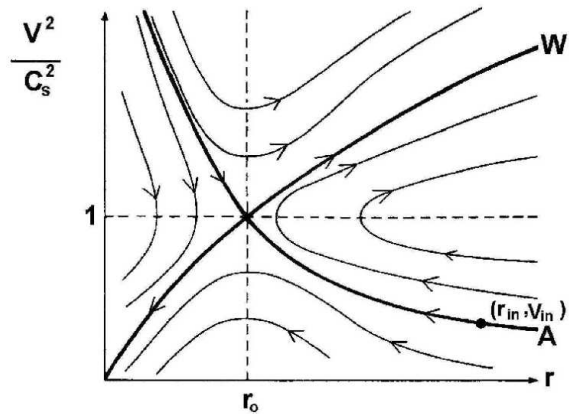}
\caption{\label{f1} \small{Stationary solutions for spherically symmetric
accretion onto a star. The bold solid curves, {\bf A} and {\bf W},
represent ``accretion" and ``wind", respectively. The fixed point is at
$r=r_0$ and $(v^2/c_s^2)=1$. A linear stability analysis indicates that
the fixed point of the flow is a saddle point. The direction of the
arrows along the curve {\bf A}, demonstrates that the transonic flow
is not physically realisable in the stationary framework.}}
\end{center}
\end{figure}

It is in principle possible to eliminate either $v$ or $\rho$ and solve
for the other variable as a function of $r$. However, adopting a slightly
different approach, it is
possible to recast Eqs.(\ref{stateuler}) and (\ref{statcon}) in a 
combined form as
\begin{equation}
\label{dvdr}
\frac{d}{dr} \left(v^2 \right) = \frac{2 v^2}{r} 
\left(\frac{2 c_s^2 - GM/r}{v^2 - c_s^2} \right) , 
\end{equation}
in which we use the speed of sound, $c_s$, to scale the flow velocity.

The various classes of solutions of Eq.(\ref{dvdr}) are shown in 
Fig.~\ref{f1} (to be presently read without the arrows).
The two dark solid curves labelled {\bf A} and {\bf W} refer to the
accretion flow and the wind flow, respectively. The meaning of  
``accretion" and ``wind" flows, which emerges from Fig.~\ref{f1} is 
quite clear. At the point $v^2 =c_s^2$, these solutions correspond to 
a finite slope, i.e. a finite value of $dv/dr$. This
implies that when $v^2=c_s^2$, we must also have $2 c_s^2=GM/r$,
implying that the intersection point is a critical point. 
The ``wind" and ``accretion" flows smoothly pass through the point 
$v^2=c_s^2=GM/2r$, with the sonic length scale having been labelled
as $r=r_0$. 

The question which now arises is that of the natural preference of 
the accreting system for a particular solution from among all the 
various possible classes of flows shown in Fig.~\ref{f1}. Linear 
stability analysis of the various stationary inflow solutions in 
real time, indicates that under the influence of a linearised 
time-dependent perturbation, all solutions appear to be stable, 
and thus it offers no clue as to the selection of any particular 
solution~\cite{pso80,gar79,td92}. To resolve this
issue, it would then be very much worthwhile to recall 
Bondi's conjecture on this
point, that the selection would be in favour of that solution, with
which is associated the least total energy. The transonic
branch satisfies that criterion, and hence is the choice. 

\section{The spherically symmetric flow as a dynamical system}
\label{sec3}

Thus far we have obtained a seemingly wholesome picture regarding
the realisability of the transonic flow. However, upon a closer  
inspection of Fig.~\ref{f1}, we find that there is a problem with it. 
For solutions which represent flows, the associated sense of direction
has been assigned by an arrow to each solution. An integration of $dv/dr$
would proceed if we start with an initial condition $v=v_{\rm{in}}$
at $r=r_{\rm{in}}$ far away from the star. For a physically realisable
flow, an initial condition infinitesimally close to a point on
the accretion line {\bf A}, would trace out a curve infinitesimally
close to {\bf A} and in the limit would correctly reproduce {\bf A},
evolving along it and passing through the critical point (sonic
point) as we integrate $dv/dr$, obtained from Euler's
equation. We will soon show that the arrows on the integration route
are as shown in Fig.~\ref{f1}. It is obvious from the direction of the
arrows here that the stationary spherically symmetric transonic accretion 
flow is not physically realisable, and closely related to this, the fixed
point is also seen to be an unstable saddle point.

To obtain any idea about the direction associated with a solution in the
steady picture, it is a matter of common knowledge in the study of dynamical 
systems~\cite{js77} that we cannot turn to Eq.(\ref{dvdr}) in its present 
form. Rather, it would be instructive to write Eq.(\ref{dvdr}) in a 
parametrised form
\begin{eqnarray}
\label{para}
\frac{d}{d \tau}\left(v^2\right) &=& 2 v^2 \left ( 2 c_s^2 - 
\frac{GM}{r} \right ) \nonumber \\
\frac{dr}{d{\tau}} &=& r \left( v^2 - c_s^2 \right) . 
\end{eqnarray}
As $\tau$, which is an arbitrary parameter, evolves ($\tau$ is
not time, since we are dealing with a stationary flow), we generate the
$v(r)$ curves. The particular curves which represent transonic flow
are the curves which pass through the fixed point at $r=r_0$, 
$v=v_0$, obtained from Eq.(\ref{dvdr}), such that 
$v_0^2=c_{s0}^2$ and $2c_{s0}^2=GM/r_0$. The
subscripted label $0$ represents physical quantities at the 
critical point. We now need to analyse the nature of the fixed
point $(r_0,v_0^2/c_{s0}^2)$. 
Writing $v^2=v_0^2+\delta v^2$
and $r=r_0 + \delta r$, and linearising in $\delta v^2$ and
$\delta r$, we find 
\begin{eqnarray}
\label{linpara}
\frac{d}{d \tau} \left(\delta v^2 \right) &=& 
2 v_0^2 \left [- \left(\gamma -1 \right)
\delta v^2 - \left( 2 \gamma -3 \right)
\frac{GM}{r_0^2} \delta r \right ] \nonumber \\
\frac{d}{d \tau} \left(\delta r \right) &=& 
r_0 \left [\frac{\gamma + 1}{2} \delta v^2
+ 2 \left(\gamma-1 \right) \frac{c_{s0}^2}{r_0} \delta r \right ] . 
\end{eqnarray}
Using solutions of the 
form $\delta v^2 \sim \exp \left(\lambda \tau \right)$
and $\delta r \sim \exp \left(\lambda \tau \right)$, the eigenvalues 
of the stability matrix implied by Eqs.(\ref{linpara}) are found to be 
\begin{equation}
\label{eigen}
\lambda = \pm c_{s0}^2 \sqrt{2 \left(5-3\gamma \right)} . 
\end{equation}
For the admissible range of $\gamma$, i.e. $1< \gamma < 5/3 $, 
the eigenvalues are real with different signs and the fixed point 
$(r_0,1)$ in the $r$ --- $(v^2/c_s^2)$ space is, therefore, identified 
as a saddle point. The arrows (characterising a saddle point) which 
are needed to make our understanding of the stationary phase portrait 
complete~\cite{js77}, are as shown in Fig.~\ref{f1}. The curves which 
we have labelled ``accretion" and ``wind" in Fig.~\ref{f1} are in 
fact now seen to be the separatrices of a dynamical system, and one 
cannot traverse the length of a separatrix upto the critical point 
in any finite range of $\tau$ values. Starting from an initial point 
to the right of the fixed point on the curve labelled {\bf A}, one 
would need an infinitely large number of steps to reach the fixed 
point, and will certainly not cross it.

To explicitly establish this result we consider 
the two parametrised equations in $\delta v^2$ and $\delta r$,
given by the Eqs.(\ref{linpara}), and using them, we write       
\begin{equation}
\label{ratio}
\frac{d \left(\delta v^2 \right)}{d \left(\delta r \right)}
= \frac{d \left(\delta v^2 \right)/d \tau}
{d \left(\delta r \right)/d \tau} . 
\end{equation}
We then integrate Eq.(\ref{ratio}) in $\delta v^2$ and $\delta r$,  
and fix the integration constant from the critical point condition,
$\delta v^2 = \delta r =0$, to obtain,
\begin{equation}
\label{veesq}
\delta v^2 = \frac{-2 c_{s0}^2}{r_0 \left(\gamma + 1 \right)}
\left[ 2 \left(\gamma -1 \right) \pm \sqrt{2 \left(5-3 \gamma \right)}
\right ] \delta r . 
\end{equation}
Using Eq.(\ref{veesq}) in the latter of the two relations given by 
Eqs.(\ref{linpara}), we get,
\begin{equation}
\label{aar}
\frac{d \left(\delta r \right)}{d \tau}= \pm c_{s0}^2
\sqrt{2 \left(5-3 \gamma \right)} \,\,\, \delta r . 
\end{equation}
We can integrate Eq.(\ref{aar}), for both roots, from an arbitrary
initial value of $\delta r =[\delta r]_{\rm{in}}$ to a point
$\delta r = \epsilon$, where $\epsilon$ is very close to the
critical point given by $\delta r =0$. We thus get 
\begin{equation}
\label{tau}
\tau = \pm \frac{1}{c_{s0}^2 \sqrt{2 \left(5-3 \gamma \right)}}
\int_{[\delta r]_{\rm{in}}}^{\epsilon} \frac{d \left(\delta r \right)}
{\delta r} = \pm \frac{1}{c_{s0}^2 \sqrt{2 \left(5-3 \gamma \right)}}
\ln \Bigg{\vert} \frac{\epsilon}{[\delta r]_{\rm{in}}} \Bigg{\vert} , 
\end{equation}
from which
it is easy for us to see that for $\epsilon \longrightarrow 0$,
$\vert \tau \vert \longrightarrow \infty$. This implies
that the critical point may be reached along either of the
separatrices, only after $\vert \tau \vert$ has become 
infinitely large. That is why the spherically symmetric flow 
cannot be realised.

It might also be noted from Fig.~\ref{f1} that within the framework of 
the stationary picture, there is another obstacle of a more practical
nature, standing in the way of the realisability of 
the transonic flow. Each of the inflow solutions --- the subsonic ones,
the transonic one and the so called ``bouncing solutions"~\cite{pso80},
is to be obtained by its own very precisely defined boundary 
condition. From among the infinitude of possibilites, the 
boundary condition that would exactly reproduce the transonic accretion
curve would have to be defined with infinite precision. And yet, even 
if that practical difficulty were to be satisfactorily addressed,
by dint of the fixed point being a saddle, the transonic inflow 
solution would still not be generated spontaneously. 

To illustrate all these points further we carry out a simple numerical
analysis. An integration of Eq.(\ref{stateuler}) with the help of the
polytropic equation of state, and its relation to the speed of sound, 
will give 
\begin{equation}
\label{integeuler}
\frac{v^2}{2} + n c_s^2 - \frac{GM}{r} = E , 
\end{equation}
in which, $n$ is the the polytropic index, given by 
$n=(\gamma -1)^{-1}$~\cite{sc39}, while 
the integration constant is fixed as $E=n{c_s}^2(\infty)$, for 
the boundary condition $v \longrightarrow 0$, 
$c_s \longrightarrow c_s(\infty)$ for $r \longrightarrow \infty $. 

Integration of the continuity equation, as given by Eq.(\ref{statcon}), 
will yield
\begin{equation}
\label{integcon}
4 \pi \rho v r^2 = \dot{m} , 
\end{equation}
in which the integration constant $\dot{m}$ (mass accretion rate) 
is given by~\cite{fkr92}
\begin{equation}
\label{emdot}
\dot{m} = \pi G^2 M^2 \frac{\rho_\infty}{c_s^3(\infty)} 
\left ( \frac{2}{5 -3 \gamma} \right )^{\left(5 - 3 \gamma \right)/
2\left ( \gamma -1 \right)} . 
\end{equation}
Combining Eqs.(\ref{integeuler}) and (\ref{integcon}), along with 
substituting $\rho$ by its dependence on $c_s$, will finally give
\begin{equation}
\label{veeaar}
\frac{v^2}{2} + n \left (\frac{\dot{\mu}}{vr^2} \right)^{1/n} 
- \frac{GM}{r} - n c_s^2(\infty) = 0
\end{equation}
in which $\dot{\mu} =(\dot{m}/4 \pi \rho_\infty){c_s}^{2n}(\infty)$.

\begin{figure}[t]
\begin{center}
\includegraphics[angle=-90,scale=0.35]{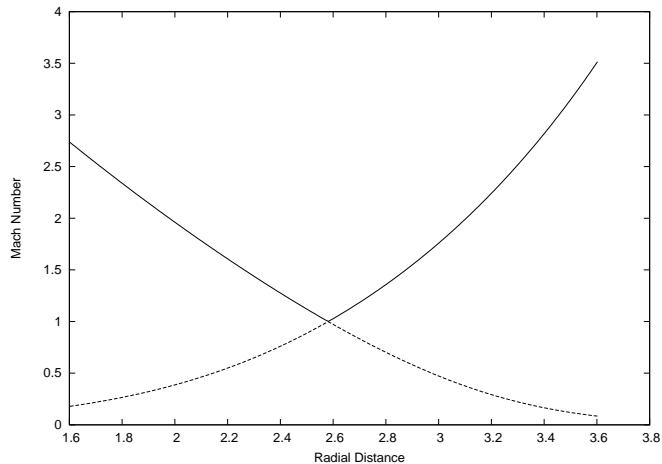}
\caption{\label{f2} \small{A plot of the Mach number, $v/c_s$, against
the radial distance $r$ (given on a logarithmic scale). The sequence
in which the data points have been numerically generated has been shown
here by the two curves.}}
\end{center}
\end{figure}

We solve Eq.(\ref{veeaar}) for $v$ numerically by the bisection method, 
using 
the values $M = M_{\odot}$, $c_s(\infty)= 10 \ {\rm{km}}~{\rm{s}}^{-1}$, 
$\rho_\infty = 10^{-21} \ {\rm{kg}}~{\rm{m}}^{-3}$ and $n=2.5$. All these
values are typical of accretion of the interstellar medium on to an
average star. Corresponding to a given boundary condition, each value of
$r$ in Eq.(\ref{veeaar}) would give a set of two real and meaningful 
solutions. The sequence of data points obtained would indicate that the 
twin solutions would look as shown in Fig.~\ref{f2} (in which the flow 
velocity has been scaled as the Mach number),  
which supports our contention about the non-realisability of the critical
solutions within the framework set up by the stationary equations alone. 

\section{Dynamic selection of separatrices : A model}
\label{sec4}

In the Section~\ref{sec3} we discussed the non-realisability of the 
stationary transonic solutions if the hydrodynamic accretion problem 
were to be studied solely in the stationary limit. The non-realisability
arises due to the fact that the stationary transonic solutions are 
actually separatrices of various classes of solutions (as the direction
of the arrows in Fig.~\ref{f1} would show) in a dynamical system. However, 
it is widely maintained that transonic solutions are indeed to be found in
a natural system, as has been seen for the case of the solar 
wind~\cite{par58,par66}, which may be treated as a spherically symmetric 
transonic outflow solution. To reconcile this observational fact with our
study of the stationary flow picture, it must be appreciated that a real
astrophysical problem is not stationary but dynamic (time-evolutionary)
in nature. Therefore we need to take into account explicit 
time-dependence of the flow variables concerned. In that case it becomes 
evident that the actual velocity profile would not only depend on 
the radial distance, $r$, but also on time, $t$. In this dynamic situation,
it is our contention that all the adverse implications regarding 
transonicity would disappear, and a transonic flow would be realised. 

\begin{figure}[t]
\begin{center}
\includegraphics[scale=0.3]{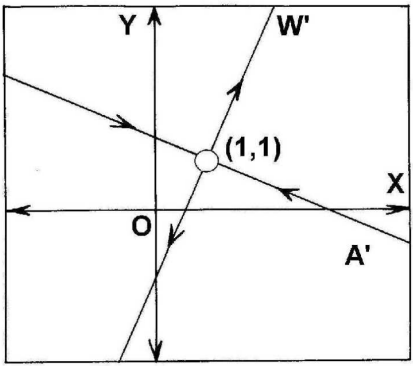}
\caption{\label{f3} \small{Integration of Eq.(\ref{statmod}) gives 
a pair of straight lines, with the integration constant fixed by 
the intersection point $(1,1)$. In the figure the lines are marked 
${\bf A}^{\prime}$ and ${\bf W}^{\prime}$. Linear stability analysis 
indicates that for ${\bf A}^{\prime}$ and ${\bf W}^{\prime}$, the
intersection point $(1,1)$ is actually a saddle point, for which the
arrows are as shown above.}}
\end{center}
\end{figure}

To assure ourselves that such indeed should be the case, we first 
consider a tractable mathematical model problem. We choose a differential
equation given by 
\begin{equation}
\label{statmod}
\frac{dy}{dx} = \frac{f(x,y)}{g(x,y)} = \frac{x+y-2}{y-x} , 
\end{equation}
whose integral can be written as 
\begin{equation}
\label{integmod}
x^2 - y^2 - 4x + 2xy = -C ,
\end{equation}
with $C$ being a constant. This is the equation of a hyperbola. If we want 
that particular solution which 
passes through the point where $f(x,y)=g(x,y)=0$, namely $x=y=1$,
then $C=2$. The curve $x^2-y^2-4x+2xy=-2$ factorises into a 
pair of straight lines : $y-x(1+\sqrt{2})+\sqrt{2}=0$ and
$y-x(1-\sqrt{2})-\sqrt{2}=0$, which are the asymptotes of the hyperbola. 
This pair is shown in Fig.~\ref{f3} as the lines 
marked ${\bf A}^{\prime}$ and ${\bf W}^{\prime}$.

We want to explore the process of drawing the line ${\bf A}^{\prime}$
from a given starting condition. On the line ${\bf A}^{\prime}$, 
we have $y=0$ at $x=2+\sqrt{2}$.
Let us begin our generating a solution with the condition $y=0$ at 
$x=2+ \sqrt{2} - \epsilon$, where $0< \epsilon \ll 1$. This starting 
condition fixes
the constant $C$ as $C=2[1+ \sqrt{2} \epsilon - \epsilon^2/2]$. Using 
this value of $C$, we can plot the curve given by Eq.(\ref{integmod}).  
For a given value of $x$, the value of $y$ is given by the relevant 
root of the quadratic equation thus obtained. The two roots are  
\begin{equation}
\label{roots}
y = x \pm \sqrt{2} \left[\left(x-1 \right)^2 + \sqrt{2} \epsilon 
- \epsilon^2/2 \right]^{1/2} . 
\end{equation}
Clearly, to satisfy $y=0$ at $x=2+ \sqrt{2}- \epsilon$, the negative
sign has to be chosen in Eq.(\ref{roots}), which will give
\begin{equation}
\label{negroot}
y = x - \sqrt{2} \left[\left(x-1 \right)^2 + \sqrt{2} \epsilon
- \epsilon^2/2 \right]^{1/2} . 
\end{equation}
At $x=0$, $y=- \sqrt{2}(1+ \sqrt{2} \epsilon - 
\epsilon^2/2)^{1/2}$, very different from $y= \sqrt{2}$,
which one gets on the line ${\bf A}^{\prime}$. In the limit of 
$\epsilon \longrightarrow 0$, one generates a part of ${\bf A}^{\prime}$
$(x \geq 1)$ and a part of ${\bf W}^{\prime}$ $(x \leq 1)$, instead of the 
whole of line ${\bf A}^{\prime}$. Another way of stating this is that the 
tracing of ${\bf A}^{\prime}$ is utmostly sensitive to initial conditions. 
If we make an error of an infinitesimal amount ${\epsilon}$ in prescribing 
the initial condition on ${\bf A}^{\prime}$, i.e. if
we prescribe $y=0$ at $x=2+ \sqrt{2}- \epsilon$ instead of $y=0$
at $x=2+ \sqrt{2}$, then the ``error"
made at $x=0$ relative to ${\bf A}^{\prime}$
is $2 \sqrt{2}$ which is ${\cal O}(1)$. An infinitesimal separation at one
point leads to a finite separation at a point a short distance away.
This is what we mean by saying that the line ${\bf A}^{\prime}$
(and similarly ${\bf W}^{\prime}$) should not be physically realised.

The clearest and most direct understanding of the difficulty is
achieved by recasting Eq.(\ref{statmod}) as a first-order autonomous
dynamical system, described by the set of differential equations
\begin{eqnarray}
\label{dynmod}
\frac{dy}{d \tau} &=& x + y -2 \nonumber \\
\frac{dx}{d \tau} &=& y - x
\end{eqnarray}
with $\tau$ being some convenient parametrisation. The fixed point
of this dynamical system is $(1,1)$, namely the point where $f(x,y)$
and $g(x,y)$ vanish simultaneously --- the point through which
${\bf A}^{\prime}$ and ${\bf W}^{\prime}$ pass. 
Linear stability analysis of this fixed point in the $\tau$ parameter
space, shows that it is a saddle point, with the eigenvalues $\lambda$
given by $\lambda = \pm \sqrt{2}$. The solutions passing through the 
critical point in this $x$ --- $y$ space can now be drawn with arrows 
and the result is as shown in Fig.~\ref{f3}. The distribution of
the arrows, characterising a saddle point~\cite{js77}, implies 
${\bf A}^{\prime}$ and ${\bf W}^{\prime}$ cannot be physically realised. 

We now investigate if these apparently non-realisable separatrices 
in the stationary limit may indeed be realised when we follow the 
evolutionary dynamics of $y$ through another variable $t$. Accordingly, 
we return to our pedagogic example of Eq.(\ref{statmod}) but
now consider $y$ as a field $y(x,t)$ with the evolution through $t$
represented by 
\begin{equation}
\label{evo}
\frac{\partial y}{\partial t}+ \left(y-x \right) \frac{\partial y}
{\partial x} = y+x-2 . 
\end{equation}
The stationary solution $y(x)$ satisfies Eq.(\ref{statmod}) and the 
discussion that follows Eq.(\ref{statmod}) is valid for $y(x)$ here. 
The stationary solutions $y(x)$ are as shown in Fig.~\ref{f3} and
the separatrices are $y(x)=x(1+ \sqrt{2})- \sqrt{2}$ and
$y(x)=-x( \sqrt{2} -1)+ \sqrt{2}$. We will now show that the dynamics 
actually preferentially selects these separatrices. 

The general solution of Eq.(\ref{evo}) can be obtained by the method 
of characteristics~\cite{ld97}. The two characteristic solutions of
Eq.(\ref{evo}) are obtained from 
\begin{equation}
\label{char}
\frac{dt}{1}=\frac{dx}{y-x}=\frac{dy}{y+x-2} . 
\end{equation}
They are
\begin{eqnarray}
\label{charsol}
y^2 -2xy- x^2 + 4x &=& C \nonumber \\
\left[x-1 \pm \frac{1}{\sqrt 2} \left(y-x \right)\right]
e^{\mp \sqrt{2}t} &=& \tilde{C}
\end{eqnarray}
with the latter having been derived by integrating the $dx/dt$ equation
with the help of the solution of the $dy/dx$ equation. The general 
solution of these two equations is given by the condition 
$C=\zeta(\tilde{C})$, where $\zeta$ is an arbitrary function, 
whose behaviour is to be determined by the initial condition at $t=0$. 

As in a physically realistic situation, we impose the condition that the
evolution is driven through a positive range of values of $t$ (``time"). 
This requirement will choose the upper sign in Eqs.(\ref{charsol}), and 
the general solution of Eq.(\ref{evo}) can then be written as 
\begin{equation}
\label{gensol}
y^2 -2xy- x^2 +4x = \zeta\left(\left[x-1 + \frac{1}{\sqrt 2} 
\left(y-x \right)\right] e^{- {\sqrt 2}t}\right) . 
\end{equation}
We give the initial condition that $y(x)=0$ at $t=0$ for all $x$. This 
leads to 
\begin{equation}
\label{incon}
\zeta \left[x\left(1- \frac{1}{\sqrt 2} \right)-1\right]=- x^2 +4x , 
\end{equation}
and gives a form for the function $\zeta$ as 
$\zeta (z)={\cal A} z^2 +{\cal B} z+ {\cal C}$, in which,  
\begin{displaymath}
\label{abc}
{\cal A}=- \frac{2}{\left({\sqrt 2}-1 \right)^2}, \; \; \; {\cal B}
=- \frac{4}{{\sqrt 2}-1} , \; \; \; {\cal C}=2 . 
\end{displaymath}
With the initial condition $y(x)=0$ at $t=0$, the solution to 
Eq.(\ref{gensol}) reads
\begin{equation}
\label{finsol}
\left[y-x \left({\sqrt 2}+1 \right)+ {\sqrt 2}\right]
\left[y+x \left({\sqrt 2}-1 \right)- {\sqrt 2}\right]
={\cal B} \phi e^{-{\sqrt 2}t}+ {\cal A} \phi^2 e^{-2{\sqrt 2}t}
\end{equation}
with
\begin{displaymath}
\label{phi}
\phi \equiv \phi (x,y)=x \left(1- \frac{1}{\sqrt 2} \right) -1
+ \frac{y}{\sqrt 2} . 
\end{displaymath}
Clearly as $t \longrightarrow \infty$, the right hand side in 
Eq.(\ref{finsol}) tends to zero and we approach one of the two 
separatrices (which were otherwise non-realisable from the stationary 
viewpoint) shown in Fig.~\ref{f3}. Of the two separatrices, the one which
will be relevant will be determined by some other imposed requirement. 
For the astrophysical flow, the two separatrices are the transonic
accretion and the wind solutions. One chooses the proper sign of the 
velocity ($v<0$ for inflows, and $v>0$ for outflows) to get the flow in 
which one is interested.

\section{A non-perturbative selection of the transonic flow}
\label{sec5}

We have seen with the help of a model problem that the dynamics makes
it possible as a matter of a mathematical principle, to select solutions
which were apparently non-realisable from the stationary perspective.
We now extend that treatment to the accretion problem here. It is 
important to understand that a real astrophysical flow itself is dynamic 
in character. This implies that explicit time-dependence of the flow 
equations would have to be taken into account. 
Having said that, it must also be said that the Eqs.(\ref{euler}) and 
(\ref{con}), which 
govern the temporal evolution of the flow, do not lend themselves
easily to a ready mathematical analysis; indeed, in the matter
of incorporating both the dynamical and the pressure effects in 
the equations, short of a direct numerical treatment, the mathematical 
problem, in Bondi's own word --- ``insuperable"~\cite{bon52} --- is 
very appropriately described. 
Therefore, to have an appreciation of the governing mechanism that 
underlies any possible selection of a transonic flow, we have to adopt
some simplifications. 

In our accretion problem we study the dynamics 
in the regime of what is understood to be the pressureless motion of a 
fluid in a gravitational field~\cite{shu} --- which is a line of 
attack that is somewhat reminiscent of the methods of Hoyle and  
Lyttleton, as Bondi has mentioned in his paper~\cite{bon52}. 
Simplification 
of the mathematical equations, however, is not the only justification
for such a prescription. A greater justification lies in the fact
that the result delivered is in conformity with, what Garlick 
calls ``the more fundamental arguments of Bondi"~\cite{gar79}, that it 
is the criterion of minimum total energy associated with a solution,
that will accord it a primacy over all the others. 

An immediate consequence of adopting dynamic equations is that
the invariance of the stationary solutions under the transformation
$v \longrightarrow -v$, is lost. As a result, we now have to separately
consider either the inflows $(v<0)$ or the outflows $(v>0)$, a 
choice that we impose upon the system at $t=0$. Euler's equation,
tailored according to our simplifying requirements, is rendered as
\begin{equation}
\label{presfree}
\frac{\partial v}{\partial t} + v \frac{\partial v}{\partial r} 
+ \frac{GM}{r^2} = 0 , 
\end{equation}
which we solve by the method of characteristics~\cite{ld97}. The 
characteristic curves are obtained from
\begin{equation}
\label{charcur}
\frac{dt}{1}= \frac{dr}{v} = \frac{dv}{-GM/r^2} . 
\end{equation}
On first solving the $dv/dr$ equation, we get
\begin{equation}
\label{dvdrchar}
\frac{v^2}{2}- \frac{GM}{r} = \frac{c^2}{2} , 
\end{equation}
with $c$ being an integration constant obtained from the spatial part 
of the characteristic equation. We use this result to solve the $dr/dt$ 
equation from Eq.(\ref{charcur}), and for $c^2 > 0$, we get
\begin{equation}
\label{drdt}
\frac{2}{c r_s}\left(vr - c^2 t \right) 
- \ln\left[\frac{r}{r_s} \left(\frac{v}{c} +1 \right)^2\right]
= {\tilde c} , 
\end{equation}
in which $\tilde{c}$ is another integration constant, and $r_s$ 
is a length scale in the system defined as $r_s= 2GM/c^2$. A similar
expression can also be written for $c^2 < 0$. 

\begin{figure}[b]
\begin{center}
\includegraphics[scale=0.4, angle = -90.0]{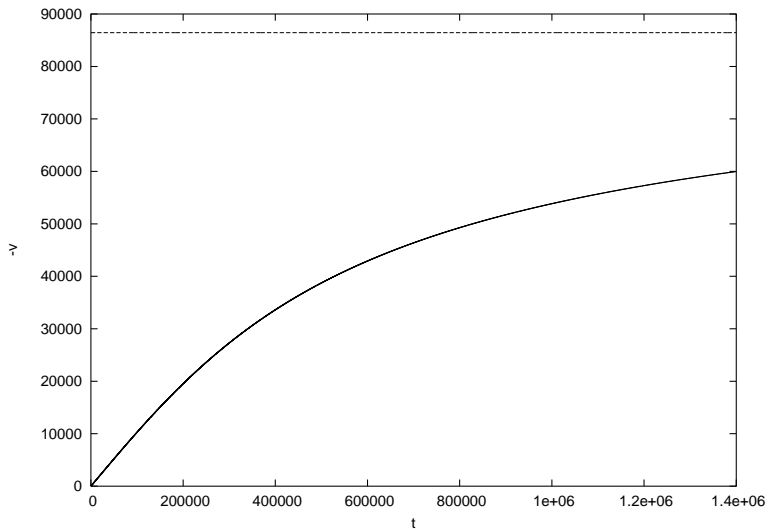}
\caption{\label{f4} \small{Evolution of the velocity field as given
by Eq.(\ref{charcur}), under the initial condition $v=0$ at $t=0$ for
all $r$. The horizontal line on top of the plot represents
a limiting value for the velocity, $\sqrt{2GM/r}$, which is being
terminally approached by $-v$, whose evolution through $t$ is
being followed at the fixed length scale, $r = 51 r_{\odot}$, with
$r_\odot$ being the radius of the accretor and with $M=M_\odot$.}}
\end{center}
\end{figure}

\begin{figure}[t]
\begin{center}
\includegraphics[scale=0.4, angle = -90.0]{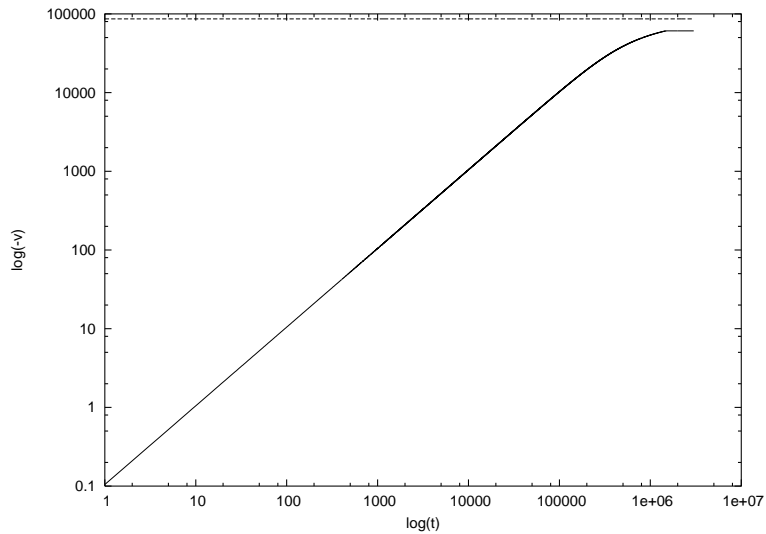}
\caption{\label{f45} \small{The slope of this logarithmic plot shows
that in the early stages of the evolution, $-v$ varies linearly with
$t$. Deviation from this linear growth sets in on time scales of
$10^5$ seconds. The horizontal line on top is the limiting value
of the velocity.}}
\end{center}
\end{figure}

A general solution of Eq.(\ref{charcur}) is given by the condition, 
${\tilde c} = \xi(c^2/2)$, with $\xi$ being an arbitrary function,
whose form is to be determined from the initial condition. We can, 
therefore, set down the general solution as 
\begin{equation}
\label{gensol2}
\frac{2}{c r_s}\left(vr - c^2 t \right)
- \ln\left[\frac{r}{r_s} \left(\frac{v}{c} +1 \right)^2\right]
= \xi \left(\frac{v^2}{2} - \frac{GM}{r} \right) , 
\end{equation}
to determine whose particular form we use 
the initial condition, $v=u_0 (r)$ at $t=0$ for all $r$, where $u_0$
is in general some initial velocity distribution over space. 
It should be easy to see that for $t \longrightarrow \infty$, 
we would get the stationary solution 
\begin{equation}
\label{statsol}
\frac{v^2}{2} - \frac{GM}{r} = 0 , 
\end{equation}
with the long-time evolutionary approach towards this stationary state 
behaving as $t^{-2/3}$. 

For a simple intuitive understanding of the physical criterion that
drives the flow towards a chosen stationary end, we set $u_0 = 0$.
This initial condition will necessitate $c^2 <0 $. 
The whole physical picture could be conceived of as one in which a  
system with a uniform velocity distribution $v=0$ everywhere, suddenly 
has a gravity mechanism switched on in its midst at $t=0$. This induces
a potential $-GM/r$ at all points in space. The system then starts 
evolving to restore itself to another stationary state, so that 
for $t \longrightarrow \infty$, the 
total energy at all points, $E=(v^2/2) - (GM/r) = 0$, remains the 
same as at $t=0$. This is evidently
the stationary solution associated with the lowest possible total
energy, and the temporal evolution selects this particular solution 
from all possible meaningful solutions.  

\begin{figure}[b]
\begin{center}
\includegraphics[scale=0.40, angle = -90.0]{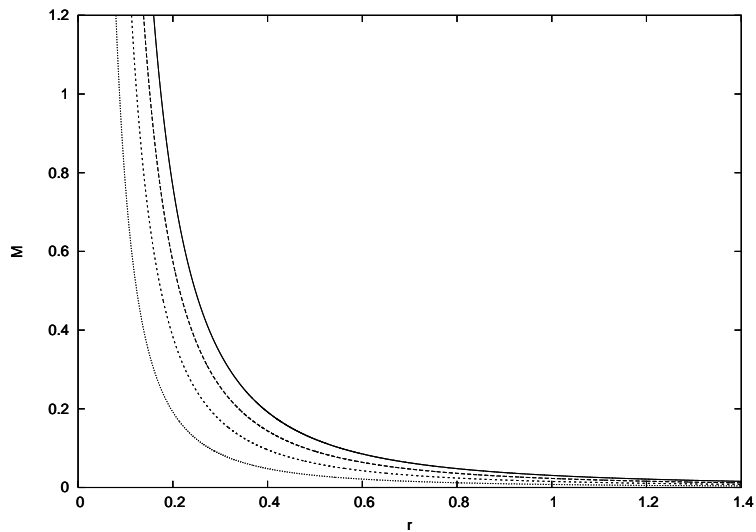}
\caption{\label{f5} \small{Evolution of the velocity field, scaled
as the Mach number, $\mathrm M$, through time, $t$. Transonicity
is clearly evident, as all curves cross the $\mathrm M = 1$ line.
Moving from left to right,
successive solutions have been shown for $t=1000$, $2000$, $3000$
and $4000$ seconds, respectively. The radial distance along the
horizontal axis has been scaled by the sonic radius, $r_0$.}}
\end{center}
\end{figure}

\begin{figure}[t]
\begin{center}
\includegraphics[scale=0.40, angle = -90.0]{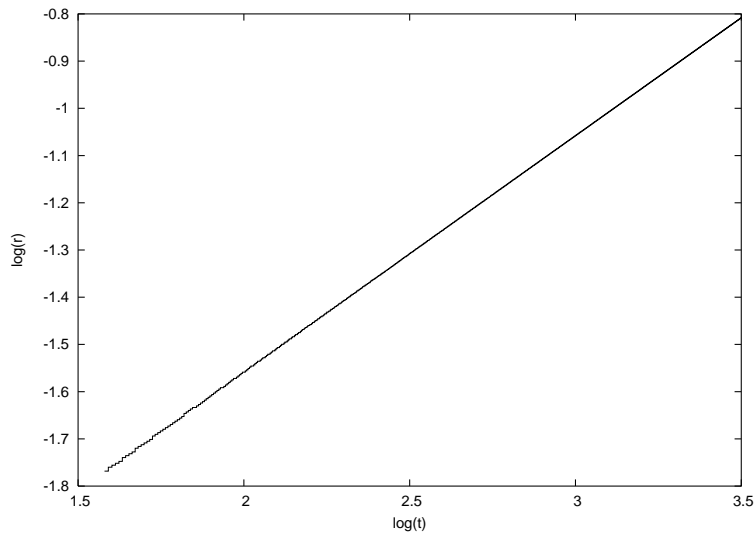}
\caption{\label{f6} \small{Progression of the sonic front in the
early stages of the temporal evolution of the flow. Once again the
radial distance along the horizontal axis has been scaled by the
sonic radius, $r_0$. The slope indicates a power dependence of
$1/2$ for $r$ versus $t$.}}
\end{center}
\end{figure}

This result has been borne out by a numerical integration 
of Eq.(\ref{charcur}) by the finite differencing technique. The 
mass of the accretor has been chosen to be $M_\odot$, while its 
radius is $r_\odot$. The evolution through time has been followed 
at a fixed length scale of $51 r_\odot$. The result of the numerical
evolution of the velocity field, $-v$ (for inflows $v$ is actually
negative), through time, $t$, has been 
plotted in Fig.~\ref{f4}. The limiting value of the velocity, as 
the evolution progresses towards the long-time limit, is evidently
$\sqrt{2GM/r}$ (with $M= M_\odot$ and $r =51 r_\odot$), 
as Eq.(\ref{statsol}) would give us to believe. 
This is what the plot in Fig.~\ref{f4} shows, as $-v$ approaches
its terminal value for $t \longrightarrow \infty$. The slope of 
the logarithmic plot of 
$-v$ against $t$ in Fig.~\ref{f45} indicates that in the early
stages of the evolution there is a linear growth of the velocity
field through time, but on later times, conspicuous deviation from 
linearity sets in.  

The foregoing argument can now be extended to understand the
dynamic selection of the transonic solution. The inclusion of the
pressure term in the dynamic equation, fixes the total energy of
the system accordingly at $t=0$. A physically realistic initial condition
should be that $v=0$ at $t=0$, for all $r$, while $\rho$ has some
uniform value. The temporal evolution of the accreting system would then
non-perturbatively select the transonic
trajectory, as it is this solution with which is associated the least
possible energy configuration. This argument is in conformity with
Bondi's assertion that it is the criterion of minimum total energy that
should make a particular solution (the transonic solution in this case),
preferred to all the others. However, this selection mechanism is
effective only through the temporal evolution of the flow.

To test this contention a numerical study has been carried out, once 
again using finite differencing, but this time using both the dynamic
equations for the velocity and the density fields, as given by 
Eqs.(\ref{euler}) and (\ref{con}). The accretor has been chosen to
have a mass, $M_\odot$, and radius, $r_\odot$. The ``ambient" conditions
are $c_s(\infty)= 10 \ {\rm{km}}~{\rm{s}}^{-1}$ and
$\rho_\infty = 10^{-21} \ {\rm{kg}}~{\rm{m}}^{-3}$, while the polytropic
index, $n=1.6$. For these values of the physical constants, transonicity
becomes apparent even at the very early stages of the evolution. The 
course of the evolution of the velocity field (scaled by the speed of 
sound), at various points of time, for a substantially representative
range of the radial distance (scaled by the sonic radius) has 
been shown in Fig.~\ref{f5}. 

The outward propagation of the sonic front, as time progresses, has
been traced in Fig.~\ref{f6}. In this logarithmic plot, what can be 
seen in the early stages of the evolution, is that the sonic front
travels through space with a $1/2$ power dependence on time. This
behaviour can be set down as $r = r_0 {\mathcal Q} t^{1/2}$, with 
the constant factor $\mathcal Q$ having been determined empirically
from dimensional considerations as
${\mathcal Q} = 4 (5 - 3 \gamma)^{-1} \sqrt{c_s^3 (\infty)/GM}$. 
This estimate tallies very closely with the numerical value obtained
from the plot in Fig.~\ref{f6}, nothwithstanding which, a cautionary 
note that has to be sounded here is that this quantitative match 
holds good for the early stages of the evolution only, and need not 
be strictly applicable for the entire span of the temporal evolution 
of the velocity field. 

\section{A perturbation equation and the metric of an acoustic black hole}
\label{sec6}

We have discussed many times in the previous sections that subjecting
the stationary solutions to a linearised time-dependent perturbation
shows that all the acceptable solutions are stable. Therefore, through 
a perturbative analysis, no clue could be had as to the special status 
of any solution. We subject this line of thinking to a closer inspection. 

To carry out a linear stability analysis Petterson et al.~\cite{pso80}
and Theuns and David~\cite{td92} have made use of a variable defined as
$f \equiv \rho v r^2$, whose stationary value is to be obtained from
the continuity equation, given by Eq.(\ref{statcon}), and this 
background value, $f_b$, is seen to be a constant that is physically 
identified with the mass flux.
In spherical symmetry, the flow variables are $v$ and $\rho$. If we 
impose small perturbations, $v^{\prime}$ and $\rho^{\prime}$, on the 
stationary background solutions, $v_b$ and $\rho_b$, we may  
derive a linearised relation for the perturbation of $f$ as 
$f^{\prime}=\left(v^{\prime} \rho_b +v_b \rho^{\prime}\right)r^2$, 
and in terms of this perturbed quantity we may then obtain a 
linearised equation of motion for the perturbation, given by 
\begin{equation}
\label{interm}
\frac{\prt}{\prt t} \left[\frac{v_b}{f_b}
\left( \frac{\prt f^{\prime}}{\prt t}\right)\right]
+ \frac{\prt}{\prt t} \left[\frac{v_b^2}{f_b}
\left( \frac{\prt f^{\prime}}{\prt r}\right)\right]
+ \frac{\prt}{\prt r} \left[\frac{v_b^2}{f_b}
\left( \frac{\prt f^{\prime}}{\prt t}\right)\right]
+ \frac{\prt}{\prt r} \left[\frac{v_b}{f_b}
\left(v_b^2 - c_{sb}^2 \right)
\frac{\prt f^{\prime}}{\prt r}\right] = 0 , 
\end{equation}
with $c_{sb}$ being the background stationary value of the speed of
sound. 

We now approach this whole question from a different perspective. It is
known that there is a close one-to-one correspondence between certain
features of black hole physics and the physics of supersonic acoustic
flows~\cite{vis98}. For an irrotational, inviscid and barotropic fluid
flow, Euler's equation may be written as
\begin{equation}
\label{baro}
\frac{\prt {\bf{v}}}{\prt t} + \frac{1}{2}{\bf{\nabla}}({\bf{v}}{\cdot}
{\bf{v}})
+ \frac{{\bf{\nabla}} P}{\rho} + {\bf{\nabla}}V = 0 , 
\end{equation}
in which $V=-GM/r$, as before.
In this situation we can represent velocity as the gradient of a 
scalar function $\psi$, i.e. ${\bf v}= -{\bf{\nabla}}\psi$. 
For the barotropic condition $\rho \equiv \rho (P)$, it is possible to
write ${\bf{\nabla}}h=({{\bf{\nabla}} P})/\rho$, upon which, from 
Eq.(\ref{baro}), we may derive the result 
\begin{equation}
\label{psi}
-\frac{\prt \psi}{\prt t} + \frac{1}{2}\left({\bf{\nabla}}\psi \right)^2 
+h+V=0 . 
\end{equation}
On a time-dependent background solution $(\rho_b, P_b, \psi_b)$ we 
impose a perturbation $(\rho^{\prime}, P^{\prime}, \psi^{\prime})$,
and together with Eq.(\ref{psi}), and the continuity equation given 
by Eq.(\ref{con}), a linear equation for the perturbation is delivered as 
\begin{equation}
\label{psipertur}
\frac{\prt}{\prt t} \left[ \frac{\rho_b}{c_{sb}^2} \left(
\frac{\prt \psi^{\prime}}{\prt t} + {\bf{v}}_b\cdot{\bf{\nabla}}
\psi^{\prime}\right) \right] + {\bf{\nabla}}\cdot \left [-\rho_b
{\bf{\nabla}} \psi^{\prime} + {\bf{v}}_b \frac{\rho_b}{c_{sb}^2} 
\left(\frac{\prt \psi^{\prime}}{\prt t} + {\bf{v}}_b\cdot{\bf{\nabla}}
\psi^{\prime}\right) \right] =0 , 
\end{equation}
in which $c_{sb}^2 = P^{\prime}/\rho^{\prime}$. This result represents
an equation of motion of an acoustic disturbance. Visser discusses this
subject~\cite{vis98} by stating that if a ``fluid is barotropic and 
inviscid, and the flow is irrotational (though possibly time-dependent)
then the equation of motion for the velocity potential describing an
acoustic disturbance is identical to the d 'Alembertian equation of 
motion for a minimally coupled massless scalar field propagating in a
$(3+1)$ - dimensional Lorentzian geometry." In other words, 
Eq.(\ref{psipertur})
may alternatively be represented by a compact formulation given as,
\begin{equation}
\label{compact}
\frac{\prt}{\prt x^{\mu}}\left (g^{\mu \nu} \frac{\prt}
{\prt x^{\nu}} \psi^{\prime} \right )=0 , 
\end{equation}
in which the Greek indices run from $0$ to $3$. 
The two expression given by Eqs.(\ref{psipertur}) and (\ref{compact}) 
are completely
equivalent~\cite{vis98}. If we consider the special case of only 
radial dependence for $g^{\mu \nu}$, then identifying $t$ as $0$ 
and $r$ as $1$ in the Greek indices, we will get from Eq.(\ref{compact}),  
\begin{equation}
\label{gees}
g^{00}=1, \; \; \;  g^{01}=g^{10}={v_b}, \; \; \; g^{11}=v_b^2- c_{sb}^2 . 
\end{equation}

Interestingly enough, the expression for $f^{\prime}$, as Eq.(\ref{interm})
shows it, can also be reduced to a similarly compact form as 
Eq.(\ref{compact}) gives for $\psi^{\prime}$, with an equivalent 
identification for an effective metric $g^{\mu \nu}$. It is easy to
compare terms between Eqs.(\ref{interm}) and (\ref{compact}) with 
$\psi^{\prime}$ replaced by $f^{\prime}$, and see that the same set
of $g^{\mu \nu}$ is obtained for $f^{\prime}$, as Eq.(\ref{gees})
shows for $\psi^{\prime}$. This closeness of form is very intriguing. 
The physics of supersonic acoustic flows
closely corresponds to many features of black hole physics. For a
black hole, infalling matter crosses the event horizon maximally, i.e.
at the greatest possible speed. By analogy the same thing may be said
of matter crossing the sonic horizon of a spherically symmetric fluid
flow, falling on to a point sink. That this fact can be appreciated
for the spherically symmetric accretion problem, through a perturbative
result, as given by Eq.(\ref{interm}), is quite remarkable. This is 
because conventional wisdom tells us that we would be quite unable 
to have any understanding of the special status of any inflow solution 
solely through a perturbative technique~\cite{pso80,gar79}. It is the
transonic solution that crosses the sonic horizon at the greatest
possible rate, and the similarity of the form between Eqs.(\ref{interm}) 
and (\ref{psipertur}) may very well be indicative of the primacy of the 
transonic solution. If such an insight were truly to be had with the 
help of the perturbation equation, then the perturbative linear stability 
analysis might not have been carried out in vain after all.

\section{Concluding remarks}
\label{sec7}

We have seen how achieving transonicity becomes distinctly easy when one 
accounts explicitly for time-dependence in the mathematical problem
of spherically symmetric accretion. Certainly the physical and 
mathematical difficulties associated with a purely static approach
to this question, disappear immediately upon involving time. But that
is not to say that all questions have been answered satisfactorily. 
For instance, a very important issue that has to be addressed in 
greater detail is the manner in which the temporal evolution drives
the velocity field towards its stationary critical (i.e. transonic)
end, especially in the long-time limit. Even in the simplified 
pressure-free regime, we have seen for
ourselves that this is not something whose answer may be given very
easily. With the involvement of the evolution of both the velocity 
and the density fields --- as it has to be for a real astrophysical
problem --- the computational difficulties will be quite staggering. 
Nevertheless, this has to be the subject of a more intensive scrutiny. 

A further intriguing issue is whether or not a perturbative analysis
in real time --- considered of not much help in understanding the 
primacy of the transonic state --- can indeed offer some insight into
questions related to transonicity. The closeness of an equation of 
motion for 
a perturbation in the accretion problem, to the metric of an acoustic
black hole, has been a beguiling revelation in this regard.

\begin{acknowledgments}
This research has made use of NASA's Astrophysics Data System.
Some numerical results presented in this article were obtained 
by using the computational facilities of Harish--Chandra Research 
Institute, Allahabad, India. 
\end{acknowledgments}

\bibliography{akrspl}

\end{document}